**Aggregation-engineered loss-tolerant strong coupling in metallic microcavities**

*Andrea Betti*, *Eleonora Cara*[*], *Giulia Serrano*, *Lorenzo Poggini*, *Alessia Valzelli*, *Natascia De Leo*, *Paolo Bartolini*, *Andrea Taschin*, *Renato Torre*, and *Alice Boschetti*[*]

A. Betti, P. Bartolini, A. Taschin, R. Torre, A. Boschetti

European Laboratory for Non-Linear Spectroscopy (LENS), Università di Firenze, Via N.Carrara 1, Sesto Fiorentino, 50019 Firenze, Italy

E. Cara, N. De Leo, A. Boschetti

Advanced Materials and Life Science Division, Istituto Nazionale di Ricerca Metrologica (INRiM), Strada delle Cacce 91, 10135 Torino, Italy

E-mail: e.cara@inrim.it, a.boschetti@inrim.it

G. Serrano

Department of Industrial Engineering (DIEF), University of Florence & INSTM RU of Florence, Via Santa Marta 3, 50139 Firenze, Italy

L. Poggini

Institute of Chemistry of Organometallic Compounds (ICCOM-CNR), Via Madonna del Piano 10, 50019 Sesto Fiorentino (FI), Italy

and

Department of Chemistry 'Ugo Schiff ' DICUS & INSTM Research Unit, University of Florence, Via della Lastruccia 3-13, 50019 Sesto Fiorentino (FI), Italy

A. Valzelli, A. Betti, R. Torre

Dipartimento di Fisica e Astronomia, Università di Firenze, Via G.Sansone 1, Sesto Fiorentino, 50019 Firenze, Italy

Andrea Taschin

Consiglio Nazionale delle Ricerche, Istituto Nazionale di Ottica, CNR-INO, Largo Fermi 6, 50125 Firenze, Italy

**Keywords**: aggregation-engineered strong coupling, organic exciton–polaritons, solution-processed metallic microcavities, molecular aggregates, excimer-mediated polariton emission, loss-tolerant polaritonics

**Abstract**

Room-temperature strong coupling in organic microcavities is usually achieved by combining high-quality optical resonators with highly ordered excitonic media, a requirement that limits scalability and processing flexibility. Here we show that this constraint can be relaxed by using molecular aggregation as a design parameter rather than treating it as a parasitic effect. We realize solution-processed Rhodamine 6G–poly(vinyl alcohol) films embedded in low-quality-factor silver Fabry–Pérot microcavities and demonstrate clear angle-resolved anticrossing with coupling energies up to 324 meV despite the large optical losses of the metallic mirrors. A two-exciton coupled-oscillator model shows that the relative weight of these species controls the collective coupling strength and can be tuned through dye loading and spin-coating conditions. In contrast, angle-resolved photoluminescence is dominated by a broad, red-shifted lower-polariton emission, consistent with relaxation through excimer-like states formed in densely packed molecular domains. These results identify molecular aggregation as a practical design lever for loss-tolerant strong coupling in wet-processed metallic cavities and suggest that ground-state aggregates and excited-state excimer-like species play distinct roles in polariton formation and emission.

## 1. Introduction

Strong light–matter coupling provides a route to engineer the optical, electronic and chemical properties of materials by hybridizing confined electromagnetic modes with molecular or solid-state excitations. When the coherent exchange of energy exceeds dissipative losses, the system enters the strong-coupling regime and forms exciton–polaritons, mixed light–matter eigenstates whose dispersion, lifetime and radiative properties can be tailored through both cavity design and material choice.[1] Organic materials are particularly attractive for room-temperature polaritonics because their large oscillator strengths and high exciton binding energies enable sizable Rabi splittings without the need for cryogenic operation. These features have stimulated intense interest in organic and hybrid polaritonic systems for low-threshold light emission,[2,3,4] nonlinear optics,[5] energy transport,[6,7] and chemical reactivity control.[8,9]

A central challenge remains the realization of strong coupling in architectures that are simultaneously robust, scalable, and compatible with simple processing. Many organic polariton platforms rely on high-quality dielectric resonators or highly ordered excitonic media to reduce optical losses and maximize coherent coupling.[10,11,12] Although these approaches have produced important demonstrations, they often require complex fabrication, stringent thickness control or molecular systems that are difficult to integrate over large areas. Metallic Fabry–Pérot microcavities offer an appealing alternative because they provide strong optical confinement in a compact and technologically simple geometry. However, the large absorption losses of metal mirrors lead to low quality factors, making the observation of robust strong coupling more challenging. Developing material strategies that can tolerate such optical losses is therefore essential for translating organic polaritonics toward scalable devices.

Supramolecular aggregation has emerged as a powerful mechanism to enhance collective light–matter interactions. In particular, J-aggregates exhibit delocalized excitonic states with large oscillator strengths and have been widely used to achieve strong coupling at room temperature.[13,14,15,16] Yet aggregation is often treated as a fixed material property or, in disordered dye-doped films, as an unwanted consequence of high molecular loading. This view is limiting. In solution-processed organic solids, intermolecular interactions can strongly reshape the excitonic landscape, generating H-like and J-like absorption features whose relative weights depend on concentration, local packing and deposition conditions. Rather than regarding this complexity as parasitic, it may be exploited as a processing-controlled degree of freedom for engineering collective coupling.

Small dye molecules, such as those performing thermally-activated delayed fluorescence, can sustain very large coupling strengths[17], even if their practical implementation is often hindered by degradation mechanisms that reduce luminescence efficiency.[18,19]

Laser dyes provide an ideal platform to explore this concept. They combine large transition dipole moments, strong visible absorption and compatibility with wet processing in polymer matrices. Rhodamine 6G (R6G), in particular, is a widely used laser dye whose photophysics is highly sensitive to concentration and local environment. In dilute solution its optical response is predominantly monomer-like, whereas in solid films high molecular loading promotes dimers and larger aggregates that broaden and shift the absorption spectrum.[20,21]

In dye-doped polymer films, increasing the dye content changes not only the density of absorbers but also the way molecules interact with one another. At sufficiently high loading, Rhodamine 6G molecules can form locally ordered domains whose optical response departs from that of isolated monomers. Depending on the relative arrangement of neighboring transition dipoles, these domains contribute absorption features with H-like or J-like character, appearing respectively on the high- and low-energy sides of the spectrum (Figure 1a). In spin-coated films, this picture is necessarily broadened by structural disorder, so that the measured response reflects an ensemble of local packing configurations rather than ideal aggregate species. Importantly, the balance between these contributions can be modified by changing the dye concentration and deposition conditions.

The same dense molecular environments can also open additional relaxation pathways after photoexcitation. While H- and J-like aggregates are ground-state assemblies that primarily reshape the absorption spectrum, excimer-like states are excited-state complexes formed dynamically between closely packed molecules. They typically produce broad, red-shifted and weakly structured emission, especially in concentrated dye films. Distinguishing these two contributions is essential in polaritonic systems: aggregation can determine the excitonic resonances that hybridize with the cavity mode, whereas excimer-like states may govern how population relaxes and radiatively recombines from the lower polariton branch.

Here we investigate exciton polaritons in planar silver microcavities embedding spin-coated Rhodamine 6G–doped poly(vinyl alcohol) (PVA) films (Figure 1b). Angle-resolved transmission measurements reveal clear anticrossing at room temperature, with coupling energies up to 324 meV despite the low-quality factor of the metallic cavity. By combining absorption spectroscopy, spectroscopic ellipsometry, transfer-matrix simulations, and coupled-oscillator modeling, we show that the polariton dispersion is governed by two aggregation-induced excitonic resonances with H-like and J-like character. Their relative contributions depend on dye loading and spin-coating conditions and determine the collective coupling strength. Angle-resolved photoluminescence, in contrast, is dominated by broad lower-polariton emission consistent with relaxation through excimer-like states formed in densely packed molecular domains. These results establish aggregation engineering as a practical route to achieve loss-tolerant organic strong coupling in simple metallic cavities and provide a scalable platform for studying polariton formation and emission in disordered molecular solids.

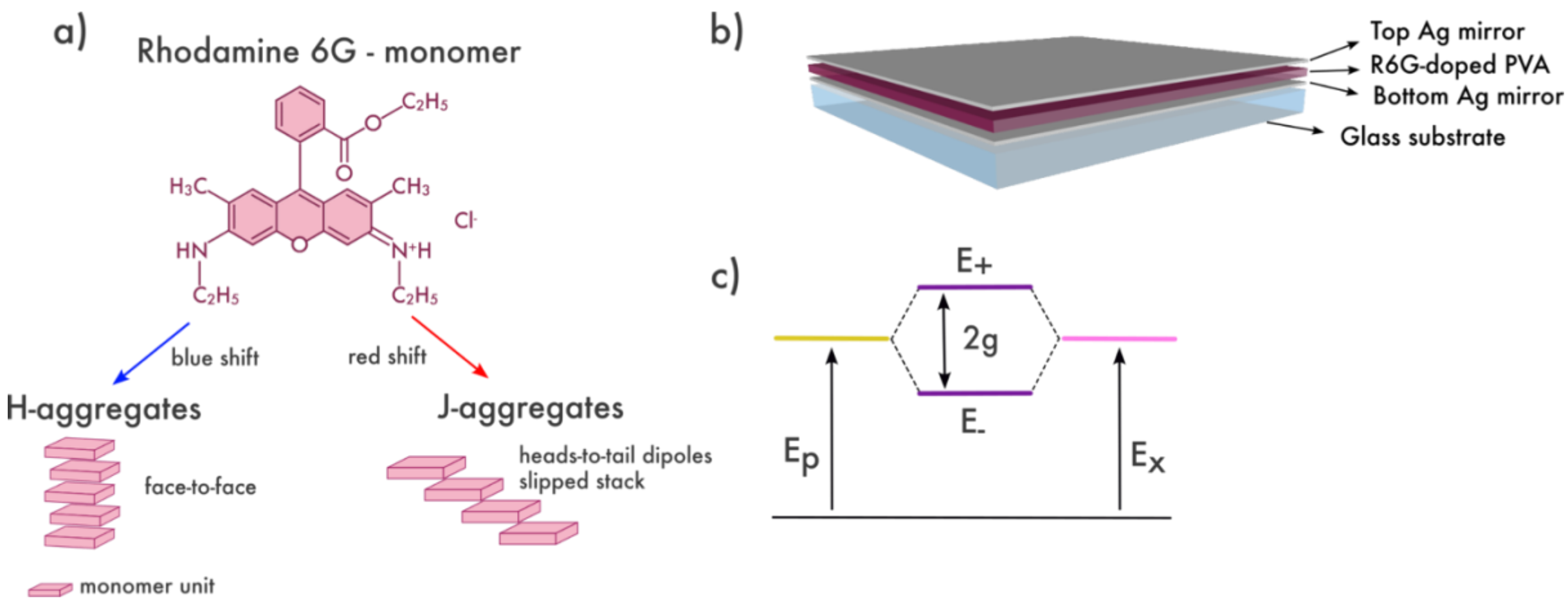

**Figure 1.** *Molecular aggregates and exciton polaritons in a planar microcavity. (a) Molecular structure of R6G and simplified representation of aggregation motifs with H-like and J-like character. H-like aggregates are associated with face-to-face dipole arrangement and a blue-shifted absorption response, whereas J-like aggregates arise from head-to-tail alignment and display red-shifted absorption. (b) Schematic representation of the planar microcavity structure, consisting of a R6G–doped PVA layer sandwiched between two semi-transparent silver mirrors on a glass substrate. (c) Energy diagram of the strong light–matter coupling regime, showing the hybridization between the cavity photon mode $E_p$ and the excitonic transition $E_x$, resulting in upper ($E_+$) and lower ($E_-$) polariton branches separated by the Rabi splitting 2g.*

### 1.1. Coupled-oscillator description

The polariton dispersion was analysed using a coupled-oscillator model, in which the confined cavity photon mode interacts coherently with one or more excitonic resonances of the absorbing layer. For a single excitonic resonance, the system can be described by the effective Hamiltonian:[22]

$$H = \begin{bmatrix} E_p & g_N \\ g_N & E_x \end{bmatrix} \qquad (1);$$

where $E_p(\theta)$ is the angle-dependent bare cavity photon energy, $E_x$ is the nondispersive exciton energy, and $g_N$ is the collective exciton–photon coupling energy. Diagonalization of this Hamiltonian gives the upper- and lower-polariton energies:

$$E_\pm = \frac{1}{2}\left(E_x + E_p(\theta)\right) \pm \frac{1}{2}\sqrt{(2g_N)^2 + (E_x - E_p(\theta))^2} \qquad (2).$$

At zero detuning, $E_p = E_x$, the energy separation between the two polariton branches corresponds to the Rabi splitting, $\Omega_r = 2g_N$ (Figure 1c).

For a single emitter the coupling energy is:

$$g_0 = |\vec{\mu} \cdot \overrightarrow{e_\lambda}|\sqrt{\frac{\hbar\omega_c}{2\,\epsilon_0\,V}} \quad (3);$$

where $V$ and $\hbar\omega_c$ are the mode volume and energy of the cavity mode, respectively, $\mu$ is the molecular transition dipole moment and $e_\lambda$ is the cavity field polarization vector.

For $N$ identical and isotropically oriented emitters $\langle|\vec{\mu} \cdot \overrightarrow{e_\lambda}|^2\rangle = \mu^2/3$,

and the collective coupling energy scales according to the Tavis–Cummings model as:[23,24]

$$g_N = \frac{g_0}{\sqrt{3}}\sqrt{N} \quad (4).$$

The angular dispersion of the bare cavity mode was modeled as:

$$E_p(\theta) = E_0/\sqrt{1-(sin(\theta)/n_{eff})^2} \quad (5),$$

where $E_0$ is the cavity-mode energy at normal incidence, $n_{eff}$ is the effective refractive index of the multilayer cavity, and $\theta$ is the external incidence angle. This expression describes the photon dispersion in the planar microcavity and was used to extract the uncoupled cavity mode from the angle-resolved transmission maps.

## 2. Results and discussion

### 2.1 Room-temperature strong coupling in low-Q silver microcavities

Planar metal–metal microcavities were fabricated by embedding a spin-coated R6G-doped PVA layer between two semi-transparent silver mirrors. The cavity geometry was selected using transfer-matrix-method calculations, with the aim of maximizing the quality factor while maintaining measurable transmission and spectral overlap between the cavity mode and the R6G absorption band. The optimized structure consists of two 40 nm silver mirrors separated by an active organic layer with a thickness of approximately 130 nm. Details of the cavity optimization for bare PVA and R6G:PVA layers are reported in the Supplementary Information (Figure S1, S2 and S3).

Angle-resolved transmission measurements were performed under transverse-electric polarization for microcavities with nominal R6G:PVA mass ratios ranging from 8% to 50%.

All samples with suitable optical thickness show two dispersive polariton branches with pronounced mode repulsion with the minimum branch separation occurring at intermediate, sample-dependent incidence angles. The upper polariton branch (UPB) appears above 2.6 eV and the lower polariton branch (LPB) below 2.2 eV, depending on concentration and cavity detuning (Figure 2a-b).

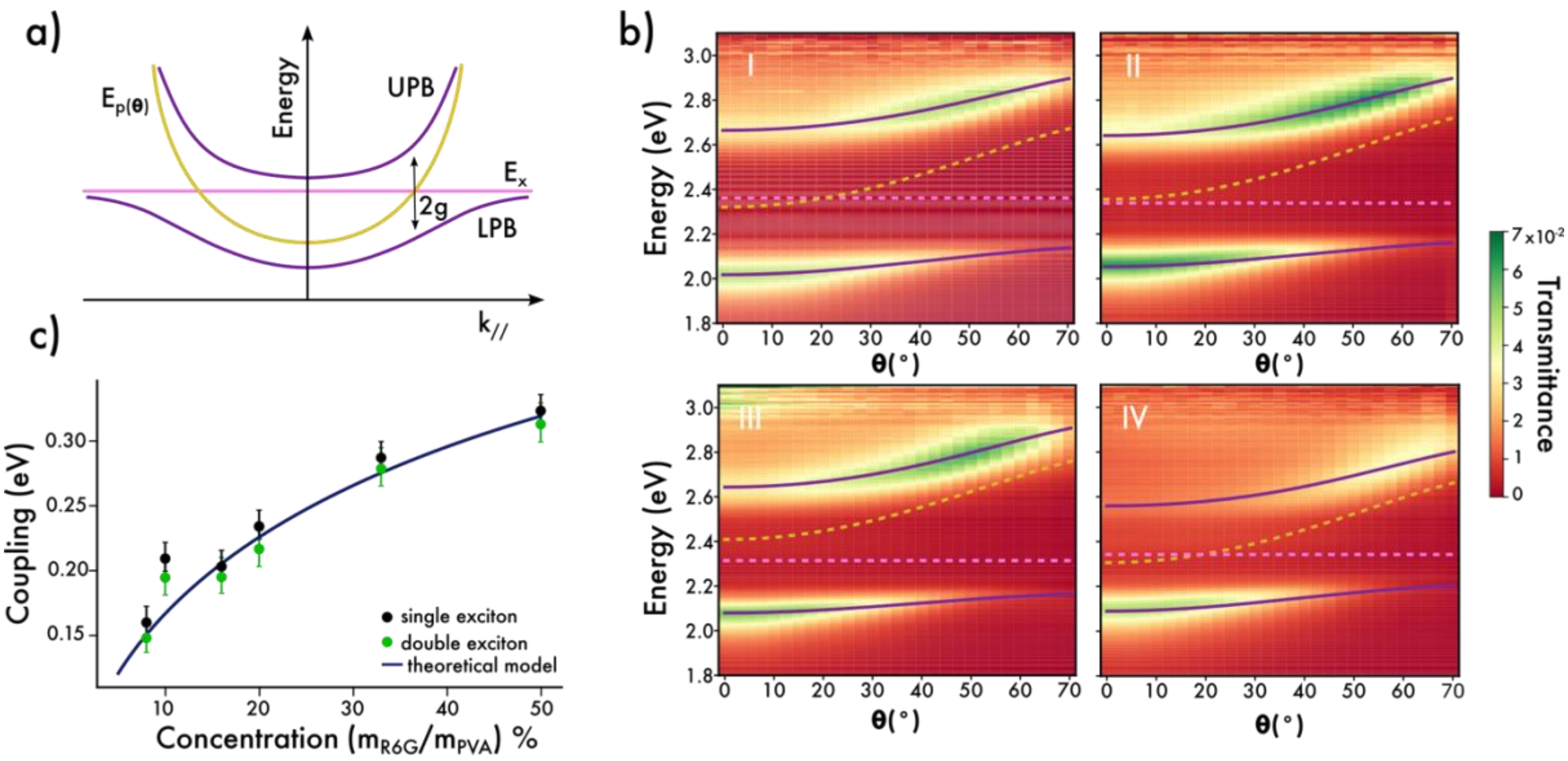

**Figure 2.** *Polariton dispersion and concentration-dependent strong coupling in R6G-based microcavities. (a) Schematic dispersion of the upper and lower polariton branches arising from the coupling between the cavity photon mode $E_p(\theta)$ and the exciton transition $E_x$. (b) Angle-resolved transmittance maps of microcavities with R6G concentrations of 50% (I), 33% (II), 25% (III), and 20% (IV). The fitted upper and lower polariton branches are shown in violet, while the uncoupled exciton and cavity-mode energies are shown in pink and yellow, respectively. (c) Extracted collective coupling energies $g_N$ as a function of dye concentration for the single- and double-exciton models. The blue line represents the theoretical concentration dependence calculated using Equation 4. The experimental data are in good agreement with the theoretical prediction.*

The measured polariton dispersions were analyzed using the single-exciton coupled-oscillator model introduced in Equation 2. The bare cavity energy at normal incidence, the collective coupling energy were treated as free parameters, while $E_x$ was constrained by the absorption spectrum of the corresponding uncapped R6G:PVA film and $n_{eff}$ and was treated as a bounded free parameter around the value extracted from the bare-cavity dispersion. For samples in which the reconstructed bare cavity and exciton modes did not reach zero detuning within the measured angular range, $g_N$ was mainly constrained by the data near their minimum fitted detuning. The good agreement between the fitted branches and the experimental transmissions confirms that the observed spectral splitting arises from hybridization between the cavity mode and the molecular transition. Within this model, the extracted collective coupling energy increases from 161 meV to 324 meV across the concentration series (Figure 2c, black dots). These values are comparable to those reported for strongly coupled supramolecular systems in dielectric cavities, despite the considerably larger optical losses of the metallic resonators.[25,26] More importantly, this trend closely follows the theoretical prediction of Equation 4 (blue line in Figure 2c), calculated without adjustable parameters assuming an isotropic orientational distribution, corresponding to an effective projected transition dipole moment of $\mu=8/\sqrt{3}$ D [27]. These results confirm that the strong-coupling regime is maintained despite the relatively low cavity quality factor, compensated by the large oscillator strength of the organic absorber. However, an additional partially resolved spectral feature emerges around 2.4 eV (Figure S4), consistent with the possible formation of a middle polariton branch. This suggests that the excitonic landscape is more complex and cannot be fully captured by the single-exciton description.

**2.2 Aggregation reshapes the excitonic landscape and controls the coupling strength**

We examined the optical properties of uncapped R6G:PVA layers, whose absorption spectra (Figure 3a) differ substantially from the spectrum of R6G in alcoholic solution (Figure S5); instead of a predominantly monomer-like absorption band, the spin-coated films display two partially overlapping resonances centered at approximately 2.46 and 2.28 eV. These features are consistent with the formation of disordered aggregates with H-like (high-energy) and J-like (low-energy) character, which reshape the absorption band. The high-energy component becomes progressively more pronounced as the nominal dye loading increases, indicating that the relative contribution of the two aggregate species changes with concentration.

Spectroscopic ellipsometry confirms this evolution (see Methods and Supplementary Info): the complex refractive indices extracted from R6G:PVA films reproduce the experimental angle-resolved transmission maps when used as input parameters in TMM simulations (Figure 3b). The extracted extinction coefficients (Figure S6) further demonstrate that the optical response of the active layers is influenced by both the dye concentration and the spin-coating conditions. In particular, increasing the dye loading or the spin speed enhances the blue-shifted absorption shoulder, consistent with a larger contribution from H-like aggregate species.

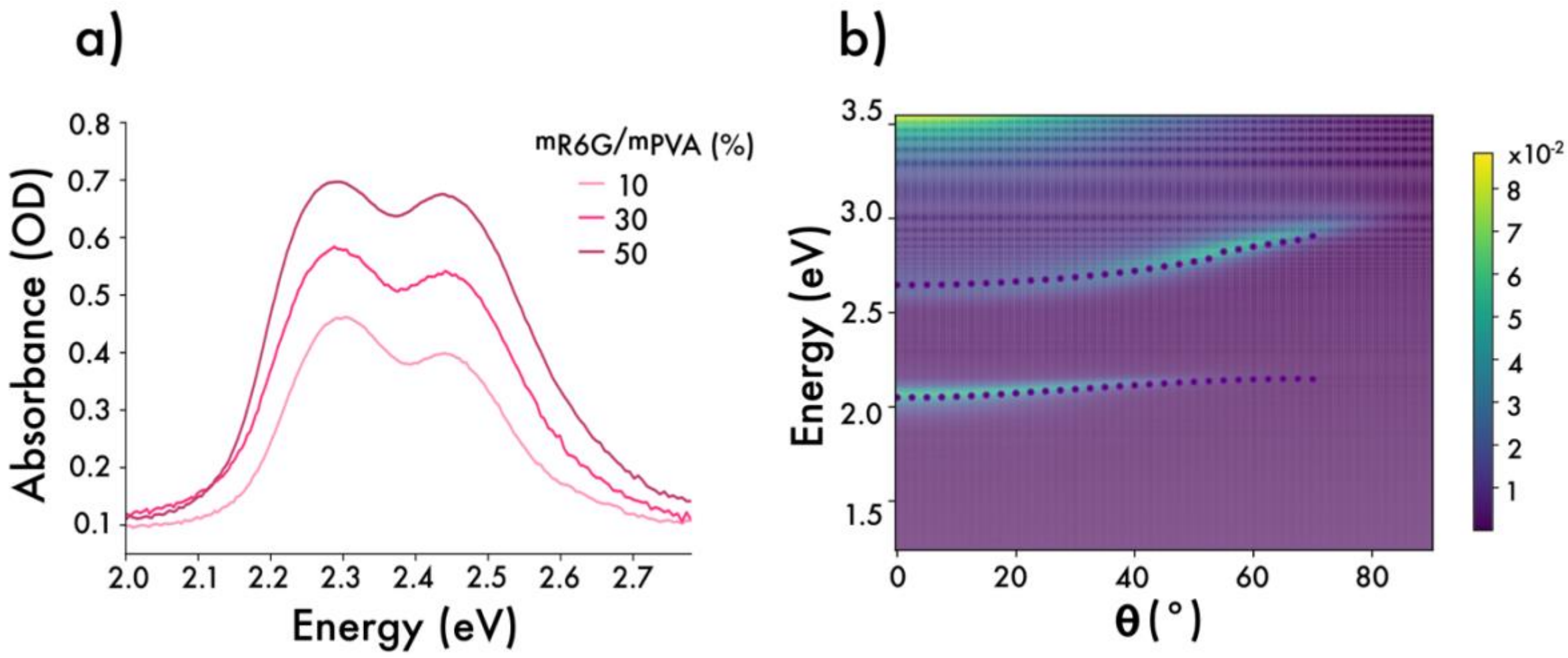

**Figure 3.** *Aggregation in uncapped R6G:PVA films and simulated cavity response. (a) Absorbance spectra of uncapped R6G:PVA layers for different nominal dye loadings ($m_{R6G}/m_{PVA}$ = 0.1, 0.3, 0.5). The spectra deviate markedly from the monomer-like response of Rhodamine 6G in solution (Figure S5) and instead display two broad resonances, consistent with the formation of disordered aggregates with H-like and J-like character. The higher-energy component becomes progressively more pronounced with increasing dye loading, indicating a growing contribution from H-like species. (b) Angle-resolved transmittance map calculated by TMM simulations using the complex refractive index extracted from spectroscopic ellipsometry of an R6G:PVA film with a nominal dye loading of 33%. The overlaid dots indicate the polariton branch energies observed experimentally for microcavities with the same R6G concentration, showing that the optical constants derived from ellipsometry accurately reproduce the measured dispersion.*

To account for the double-band absorption, we introduced a two-exciton model for the polariton dispersions (Figure 4a and Figure S7). The fitted bare cavity energies at normal incidence are comparable to those obtained using the single-exciton model (Figure S8a). The coupling energies extracted for the two aggregate-related resonances are reported as a function of dye concentration and spin speed in panels b) and c) of Figure 4. The low-energy resonance shows an initial increase in coupling up to a nominal dye loading of 33%, followed by a tendency toward saturation. In contrast, the high-energy resonance displays a monotonic increase over the same concentration range. The corresponding coupling energies vary from approximately 144 to 211 meV for the low-energy resonance and from 37 to 232 meV for the high-energy resonance. This behavior is consistent with the absorption measurements: as the dye loading increases, the high-energy H-like contribution grows, while the low-energy contribution tends to saturate.

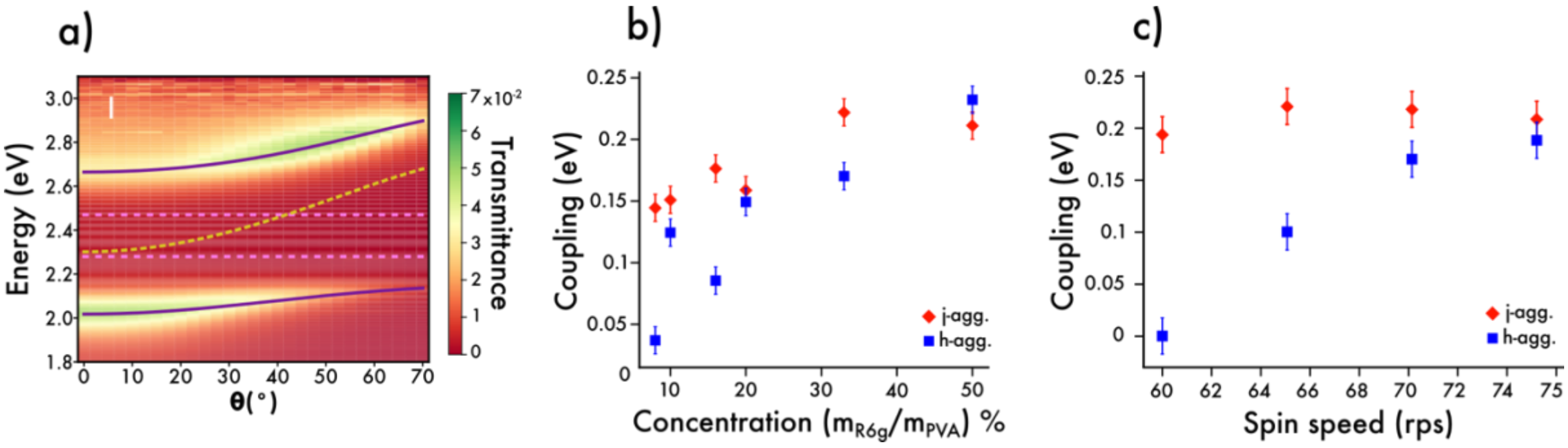

**Figure 4.** *Polariton dispersion and coupling in the double-exciton model. (a) Angle-resolved transmittance map for the 50% R6G:PVA microcavity, with the upper and lower polariton branches extracted from the double-exciton model shown in violet, the uncoupled H-like and J-like excitons and cavity-mode energies are shown in pink and yellow, respectively. (b) Collective coupling energies associated with the high-energy H-like contribution (blue) and the low-energy J-like contribution (red), extracted from the double-exciton fits of microcavity samples with comparable optical thickness, as a function of dye concentration, and (c) as a function of spin-coating speed at a fixed dye concentration of 33%.*

The effective coupling defined as the root-sum-square of the two individual coupling strengths, follows the same overall trend as the single-exciton coupling (Figure 2c), showing that both descriptions capture the total strength of the light-matter interaction. The two-exciton model, however, separates the contributions of the two aggregate-related resonances.

To investigate the effect of fabrication conditions, we analyzed microcavities containing R6G:PVA films deposited at spin-coating speeds ranging from 60 to 75 rps, while keeping the nominal dye concentration fixed at 33%. For a given solution viscosity, increasing the spin speed would generally be expected to reduce the film thickness, leading to a blue-shift of the cavity-mode resonance. However, Figure S8 shows the opposite trend. This behavior may be related to changes in the film-formation dynamics at higher spin speeds. In particular, faster solvent evaporation may lead to a more rapid increase in solution viscosity during spinning, thereby limiting lateral flow and potentially resulting in a thicker final layer.

By plotting $g_N$ as a function of the spin speed (Figure 4c), we observe a clear increasing trend, as for dye concentration. Although $g_N$ is expected to be independent of cavity thickness, the observed variation may instead reflect changes in the relative abundance of monomeric and aggregated species, or an improvement in the homogeneity of the deposited layer.

By separating $g_N$ into the relative contributions of the two-exciton model, we observe that the low-energy exciton exhibits a saturating trend with increasing spin speed, similarly to what is observed with increasing dye concentration. In contrast, the contribution of the high-energy exciton increases, despite the cavity mode energy (Figure S8b) being red-shifted at higher spin speeds. This demonstrates that the formation of aggregates with H-like character is also promoted through high rotation speed. Because the transition dipole moments associated with the two species are different, this results in a corresponding variation in the effective coupling (Figure 4c).

The emission spectra of the dye-doped layers show a similar profile across all concentrations and excitation wavelengths (Figure 5a and Figure S9), with a maximum around 2.06 eV and a broad band extending down to 1.6 eV, including a shoulder near 1.90 eV. Excitation spectra monitored at different emission energies contain contributions from both the high- and low-energy absorption regions, with a stronger contribution from the high-energy band. Since both excitonic resonances contribute to the same emission shape, this would suggest either one aggregate species or energy transfer occurring among the two species. However, the different relative intensities observed in absorption and excitation spectra supports the presence of at least two aggregate species with significant spectral overlap.

Previous studies of R6G thin films report predominantly monomeric behavior, with spectral features associated with the $S_1$ state and its vibronic progression. At higher concentrations, however, an additional high-energy absorption band emerges, consistent with the formation of H-type dimers, typically associated with a parallel stacking of the xanthene planes.[28]
The low photoluminescence quantum yield measured in our films (~1%) is consistent with strong aggregation-induced quenching, although it cannot by itself identify the specific aggregate geometry. While H-like aggregation can therefore account for the high-energy absorption feature and the reduced emission efficiency, it does not readily explain the broad and strongly red-shifted photoluminescence observed in our films. This emission could instead be consistent with the formation of excimer-like states in densely packed molecular regions. At high dye loading, structural relaxation following photoexcitation may stabilize such intermolecular excited states, giving rise to the broad, red-shifted emission characteristic of a structurally disordered local environment.[20]

It is therefore important to distinguish between aggregate and excimer states, as they originate from fundamentally different mechanisms. Aggregates are ground-state molecular assemblies, whose intermolecular coupling gives rise to modified absorption features and, in the case of H-like configurations, typically weak or quenched emission. In contrast, excimers are excited-state complexes formed dynamically after photoexcitation, producing broad, red-shifted emission due to structural relaxation in the excited state. In densely packed R6G films, the close molecular packing associated with aggregation can facilitate excimer formation upon excitation.

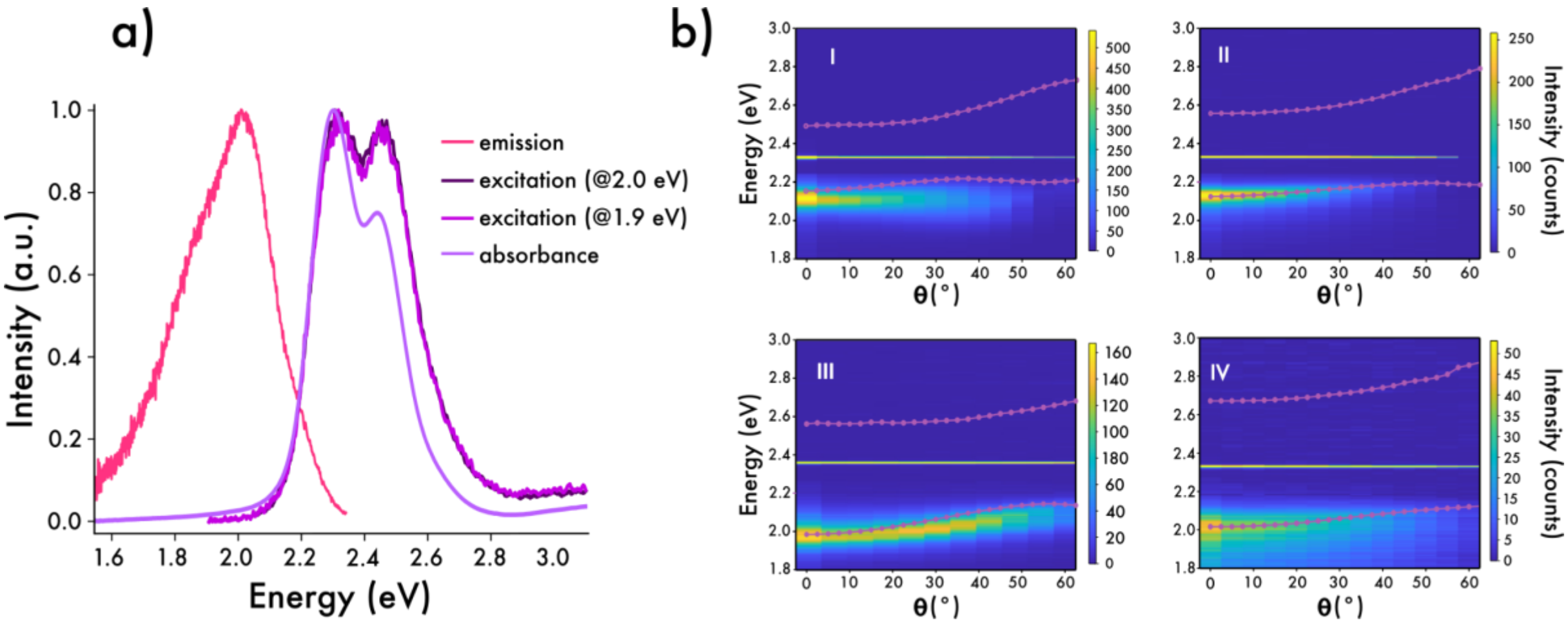

**Figure 5.** *Absorbance, excitation, and angle-resolved emission in R6G-based microcavities. (a) Absorption, emission, and excitation measurements of a R6G:PVA film at 10% concentration. Excitation measurements are collected at emission energies of 1.9 and 2.0 eV, respectively, at the maximum and at the shoulder of the emission spectrum, whose similar shape suggests that both the high- and low-energy absorption regions populate the same broad emissive manifold, either directly or through energy transfer. (b) Angle-resolved emission of 4 cavities with different concentrations 8% (I), 16% (II), 20% (III), and 50% (IV). The dots are the fitted centers of the UPB and LPB obtained by the angle-resolved transmittance measurements. The peak appearing at 532 nm is the CW laser excitation. Only emission associated with the lower polariton branch is observed because the excitation energy lies below the UPB.*

Angle-resolved photoluminescence was measured by exciting the microcavities with a CW laser at 532 nm (Figure 5b). Nearly all samples show radiative recombination from the LPB, while emission from the UPB is not observed because the excitation energy lies below the upper branch. At large emission angles, some samples exhibit weak spectral confinement and an almost angle-independent emission contribution (panel I of Figure 5b), consistent with residual uncoupled emission in spectral regions where the overlap with the LPB is reduced. The broad, red-shifted photoluminescence predominantly associated to excimer-like states may act as a relaxation reservoir that feeds the LPB when sufficient spectral overlap is present. Thus, while the polariton dispersion observed in transmission is determined by aggregate-related excitonic resonances, the red-shifted emission is more consistently associated with relaxation through excimer-like states. This distinction highlights the different roles played by ground-state aggregation in polariton formation and by excited-state intermolecular interactions in the population of the LPB.

## 3. Conclusions

In summary, we have demonstrated robust room-temperature strong light–matter coupling in a low-Q silver-based planar microcavity embedding wet-processed Rhodamine 6G in a PVA matrix. Clear anticrossing behavior and coupling energies up to 324 meV were observed, highlighting the remarkable loss tolerance of organic excitonic systems with large oscillator strengths. The experimental coupling strengths agree quantitatively with the Tavis–Cummings model, confirming the collective origin of the interaction. A central result of this work is that the coupling strength is not governed solely by dye concentration but is strongly influenced by molecular aggregation, which can be controlled through both concentration and manufacturing parameters. Spectroscopic ellipsometry and stationary spectroscopic measurements support the presence of absorption contributions with H-like and J-like character, whose relative populations evolve with fabrication conditions. The two-exciton coupled-oscillator model indicates that the coupling associated with the high-energy H-like contribution increases at higher dye loading and spin speed, becoming comparable to the low-energy contribution. These results show that processing-induced changes in molecular aggregation modify the effective oscillator strength available for coupling and provide an additional means of tuning polaritonic interactions without modifying the cavity geometry.

Interestingly, while strong coupling is established from the angle-resolved transmission measurements, the emission response reveals a more complex scenario. The photoluminescence becomes strongly red-shifted and broadened at high dye concentrations, consistent with emission from excimer-like states formed in densely packed R6G films. Angle-resolved emission is consistent with population of the lower polariton branch from these states when sufficient spectral overlap is present. This suggests that, under high-loading conditions, excimer-like states provide an important relaxation pathway toward the lower polariton branch, rather than the emission arising solely from isolated monomers or ground-state aggregates. The coexistence of aggregate-based strong coupling in absorption and excimer-assisted polariton emission suggests a multilevel relaxation landscape, where energy transfer between molecular species plays a key role in polariton population.

Our results show that strong coupling and polariton emission can be achieved in technologically simple metallic microcavities fabricated entirely by scalable deposition techniques. The demonstrated sensitivity of the coupling strength to aggregation and processing parameters introduces an additional design degree of freedom for organic polaritonic devices. This platform therefore offers a practical route toward solution-processed, loss-tolerant polaritonic light sources and provides a model system to study aggregation- and excimer-mediated polariton physics in disordered molecular solids.

## 4. Experimental Methods

*Cavity design*: A plane parallel cavity was selected due to the fabrication simplicity; the exciton energy and the EM confinement have driven its design. To determine the correct cavity dimensions, a customized Python-based routine for simulating the light propagation inside a multilayer film is used, the TMM, which includes the effect of the multiple internal reflections. The constraints are a measurable transmittance of the cavity mode, i.e. greater than 1 %, the resonance energy overlapped with the exciton energies, and a Q-factor as high as possible. The TMM permits the calculation of the dispersion of the cavity mode, which is used to tune the energy gap between the confined EM field and the dye.

*Cavity fabrication*: The microcavities were fabricated on Corning Eagle XG glass substrates (thickness < 1.1 mm). The substrates were initially cleaned by an ultrasonic bath in acetone and subsequently in isopropyl alcohol. The bottom and top mirror layers, schematized in Figure 1b, were each deposited in 40 nm-thick layers by electron-beam evaporation using a Temescal FC-2000 system (Ferrotec). Preliminary microcavity samples, with the same metal thickness, are realized using a Highly Modular Physical Vapor Deposition System (Korvus Technology), which operates via magnetron sputtering, reaching a vacuum up to $10^{-7}$ mbar and a growth rate of around 0.3 Å/s. The active layer in between the mirrors was obtained by preparing a solution of PVA in deionized water with a concentration of 40 mg/ml and mixing it in different proportions with a solution of R6G in ethanol with a concentration of 10 mg/ml. The mixing proportions of the R6G and PVA dispersions were set to 8%, 10%, 16%, 20%, 25%, 33%, and 50%, calculated as the dry-mass ratio of R6G to PVA. The realization protocol of the solutions had been optimized during the cavity fabrication to avoid cluster formations, which decrease the cavity optical quality. A fixed volume of 150 μL of the mixture was spotted on the glass substrate after the bottom mirror coating and spread on the surface by spin coating. An optimized spin-coating recipe consisted of two steps and was performed with a programmable spin coater (WS-400B-6NPP/LITE Laurell Technologies). The first step was performed at varying speeds and accelerations lasting for 170 s. The spinning speed was set to 60 rps, 65 rps, 70 rps or 75 rps to achieve different thicknesses in the deposition process, while the acceleration was set accordingly to have a 20 s acceleration ramp and 150 s of constant spinning conditions. The second consecutive step was kept the same for all samples and included a deceleration down to 20 rps for a 20 s ramp and spinning at this constant speed for another 20 s. The mirrors' deposition and active layer preparation were conducted in ISO6 and ISO7 clean-room laboratories. The microcavity samples were stored under low vacuum to avoid oxidation of the silver coating.

*Spin coating calibration*: To calibrate the spin-coated thicknesses, we fabricated more than thirty cavities with 15 nm of gold deposition and eleven different spin speed parameters. These were investigated using optical transmittance measurements to find the resonance wavelength and, using the TMM simulations, we related the resonance to the polymer thickness. Using these measures, we reached the spin speed needed to obtain the desired width. To define the thickness variability of the polymer layer in the microcavity due to the spin coating, we fabricated three different cavities at each spinning speed, finding that the thickness variability decreases with increasing spinning speed and can be kept below 20 nm for all parameters.

*Spectroscopic ellipsometry (SE)*: SE characterization was performed to determine the thickness and optical constants (n,k) of the spinned active layer with J.A. Woollam α-SE ellipsometer at 70° incident angle in the wavelength range from 300 nm to 900 nm. At the same angle, the reflected beam was detected to measure the relative change in the amplitude $\Psi$ and phase $\Delta$ between the parallel and perpendicular polarized light components. The measurements were performed on the glass substrates covered with 40 nm of Ag and spin-coated with the R6G-doped PVA layer.  The samples were modeled with CompleteEASE proprietary software, selecting a Cauchy glass substrate and a silver layer with a fixed thickness. On top of these, a B-spline layer was added to fit the $\Psi$ and $\Delta$ curves. The B-spline starting material was selected to be a Cauchy layer with optical constants matching a PVA layer, n = 1.460 and k = 0.05. The reference values were taken from 'refractiveindex.info' database[29]. The B-spline nodes resolution was set to 0.05 eV to match the absorption feature of the film, rapidly changing in the visible wavelength range due to the presence of R6G, absorbing at 550 nm, while PVA is transparent and has no absorption in the visible range. The relative permittivity $\varepsilon 2$ is forced to positive values to avoid having non-physical fitting results. To describe complex behaviors in the absorption spectrum ($\varepsilon 2$) and, at the same time, provide a physically correct dispersion spectrum ($\varepsilon 1$), Kramer-Kronig-consistent basis functions were used[30]. The B-spline model that best fitted the experimental SE data was then parametrized using Tauc-Lorentz and Gaussian oscillators using physical parameters to describe the $\varepsilon 2$ (k) curve first and then the $\varepsilon 1$ (n) curve. The thickness was then recalculated with a more accurate fitting of the $\Psi$ and $\Delta$ curves.

*Transmittance and Fluorescence measurements*: Angle-resolved transmittance measurements were performed by illuminating the samples with an in-fiber deuterium–tungsten halogen light source, DH-2000-BAL Ocean Optics, coupled to a 200 µm-core fiber. The light was collimated using a 10× microscope objective with a numerical aperture of 0.25. A telescope was used to reduce the beam spot size to approximately 3 mm on the sample, and a dichroic film polarizer was used to select TE polarization. The light transmitted through the sample was collected by a 50 mm converging lens and coupled into a 600 µm-core fiber connected to an Ocean Optics Red Tide USB650 spectrometer. The samples were mounted on a goniometer, which varied the incidence angle of the white-light beam on the cavity surface, enabling angle-resolved transmittance measurements. The same setup was used for fluorescence measurements, replacing the white-light source with a continuous-wave laser operating at 532 nm (Coherent Verdi V5).

**Acknowledgements and funding**

Part of this work was carried out at Nanofacility Piemonte, INRiM, a laboratory supported by the Compagnia di San Paolo Foundation, and at QR Laboratories, INRiM's micro- and nanofabrication facility. Additional activities were performed at the PiQuET research infrastructure (Piemonte Quantum Enabling Technologies) at INRiM, supported by Regione Piemonte. The authors acknowledge the technical support and access to facilities provided by the Interdepartmental Research Unit "MatchLab" of the University of Florence. Financial support was provided by IPHOQS – Integrated Infrastructure Initiative in Photonic and Quantum Sciences (IR0000016, CUP B53C22001750006), the CNR-FOELENS project, and the European Union's Horizon Europe Research and Innovation Programme through the EIC Pathfinder project APACE (Grant Agreement No. 101161312).

**Conflict of interest**

The authors declare no competing interests

**Data Availability**

The data that support the findings of this study are available from the corresponding author upon reasonable request.

# Supporting Information

**Aggregation-engineered loss-tolerant strong coupling in metallic microcavities**

*Andrea Betti, Eleonora Cara*, Giulia Serrano, Lorenzo Poggini, Alessia Valzelli, Natascia De Leo, Paolo Bartolini, Andrea Taschin, Renato Torre, and Alice Boschetti**

A. Betti, P. Bartolini, A. Taschin, R. Torre, A. Boschetti
European Laboratory for Non-Linear Spectroscopy (LENS), Università di Firenze, Via N.Carrara 1, Sesto Fiorentino, 50019 Firenze, Italy

E. Cara, N. De Leo, A. Boschetti
Advanced Materials and Life Science Division, Istituto Nazionale di Ricerca Metrologica (INRiM), Strada delle Cacce 91, 10135 Torino, Italy
E-mail: e.cara@inrim.it, a.boschetti@inrim.it

G. Serrano
Department of Industrial Engineering (DIEF), University of Florence & INSTM RU of Florence, Via Santa Marta 3, 50139 Firenze, Italy

L. Poggini
Institute of Chemistry of Organometallic Compounds (ICCOM-CNR), Via Madonna del Piano 10, 50019 Sesto Fiorentino (FI), Italy
and
Department of Chemistry 'Ugo Schiff ' DICUS & INSTM Research Unit, University of Florence, Via della Lastruccia 3-13, 50019 Sesto Fiorentino (FI), Italy

A. Valzelli, A. Betti, R. Torre
Dipartimento di Fisica e Astronomia, Università di Firenze, Via G.Sansone 1, Sesto Fiorentino, 50019 Firenze, Italy

Andrea Taschin
Consiglio Nazionale delle Ricerche, Istituto Nazionale di Ottica, CNR-INO, Largo Fermi 6, 50125 Firenze, Italy

**TMM Simulations**

Transfer-matrix-method simulations were performed to design the planar silver microcavity and to identify a cavity resonance spectrally overlapping with the absorption band of Rhodamine 6G (R6G). The thickness of the inner polymer layer was optimized while keeping both silver mirrors fixed at 100 nm. This preliminary analysis indicated that a polymer thickness of approximately 130 nm is required to obtain a cavity mode in the spectral region of interest. Subsequently, the thicknesses of the two silver mirrors were varied from 10 to 100 nm in order to evaluate their effect on the cavity quality factor and optical transmission. A mirror thickness of 40 nm was selected as a compromise between a sufficiently high quality factor, approximately Q ≈ 50, and a transmittance of approximately 40%. The angular transmittance map for TE polarization and the normal-incidence transmission spectrum of the optimized cavity design are reported in Figure S1.

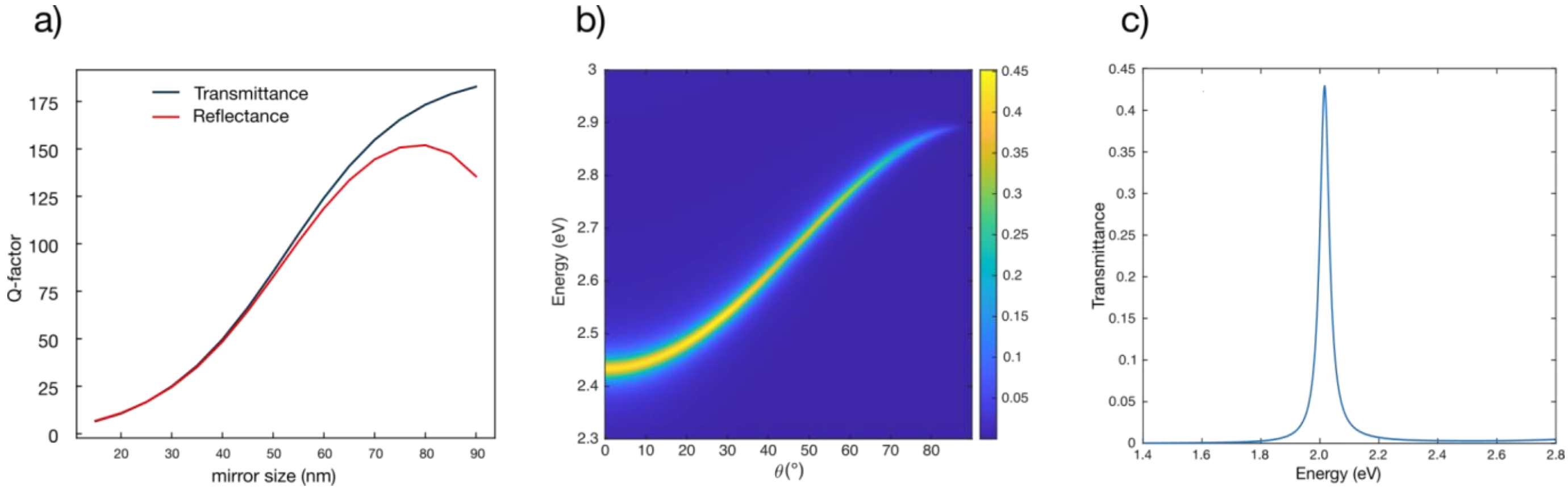

**Figure S1:** *(a) Simulated dependence of the Q-factor of a symmetric silver microcavity containing a 130 nm-thick PVA layer as a function of silver mirror thickness. (b) Simulated angle-resolved transmittance map of a symmetric microcavity with 40 nm-thick silver mirrors and a 130 nm-thick PVA layer. (c) Normal-incidence transmission spectrum of the same microcavity shown in panel (b).*

For the dye-doped microcavities, the inner polymer layer was modeled using an effective complex refractive index obtained as the weighted average of the refractive indices of R6G and PVA. The weights were defined according to the relative dry volumes of R6G and PVA. Given the similar densities of the two dry components, the volume fractions differ only slightly from the corresponding mass fractions and were therefore used as an equivalent estimate of the R6G/PVA loading. For a 130 nm-thick active layer, polariton splitting was expected at incidence angles between 30° and 40°. The simulations indicate that, for a 130 nm-thick active layer, an R6G/PVA mass ratio of approximately 10% is required to achieve a resolvable polariton splitting.

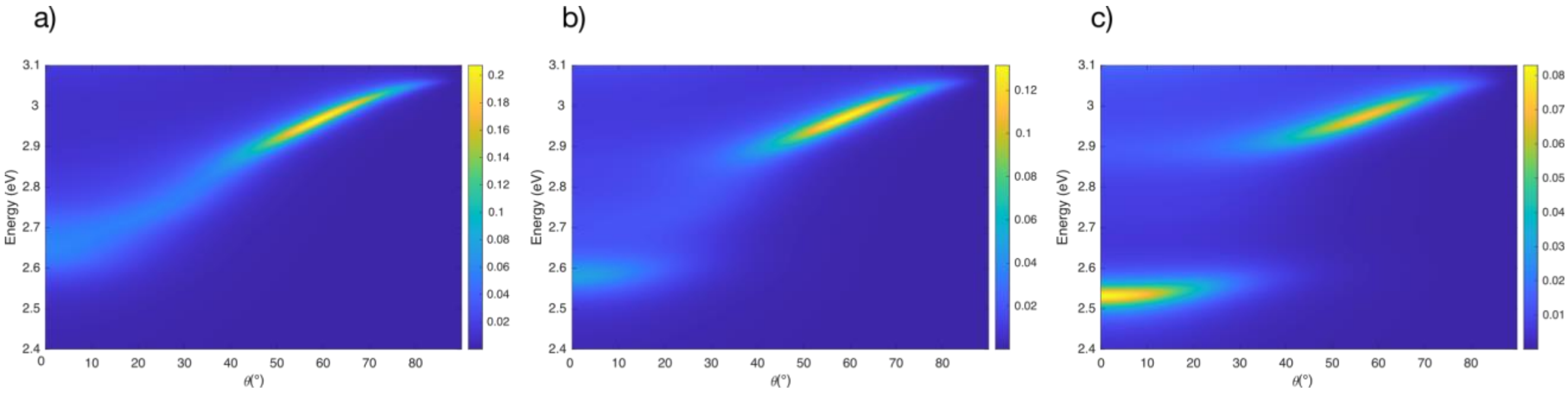

**Figure S2:** *Simulated optical response of silver microcavities containing R6G-doped PVA layers with different dye loading levels a) 5%, b) 10% and c) 20%. The silver mirror thickness was fixed at 40 nm, while the R6G:PVA layer thickness was set to 130 nm. The reported percentages correspond to the R6G/PVA mass ratio.*

**Experimental measurements of angular resolved transmittances**

The angular transmittance maps of an undoped PVA silver microcavity and of a dye-doped PVA layer on a glass substrate are shown in Figure S3. The bare cavity displays the characteristic angular dispersion of a single optical mode. The measured linewidth of this mode is 148 meV, (Q = 16), which is significantly broader than predicted by the simulations. This discrepancy is likely due to additional optical losses arising from mirror absorption, surface roughness, and polymer-thickness nonuniformity introduced during the spin-coating process, which are not fully captured by the idealized simulations. In contrast, the dye-doped PVA layer on glass shows a substantially angle-independent optical response. Angle-resolved transmission spectra of three representative R6G:PVA microcavities are reported in Figure S4. In addition to the main upper- and lower-polariton resonances, a partially resolved feature emerges around 2.4 eV. This feature is not fully described by a single-exciton model and could be consistent with the emergence of a weakly resolved middle polariton branch.

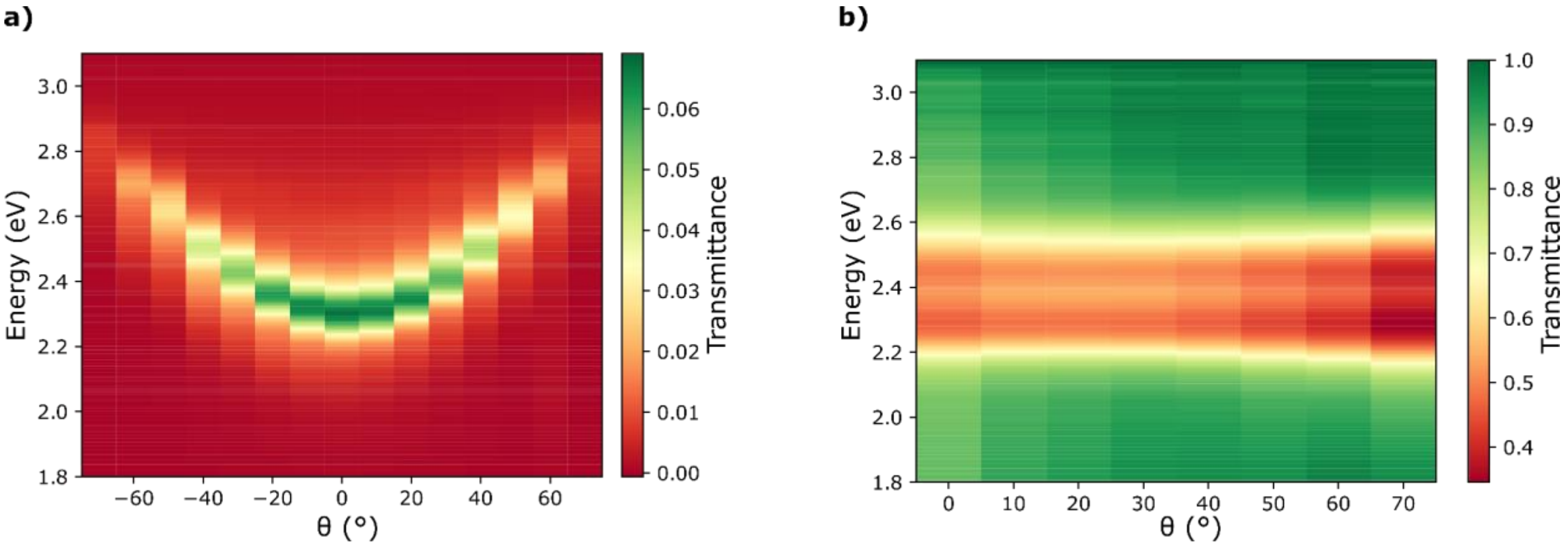

**Figure S3:** *Angle-resolved transmittance measurements of: (a) a silver-based planar microcavity containing an undoped PVA layer; (b) a thin R6G-doped PVA layer with 33% dye loading, spin-coated on a glass substrate.*

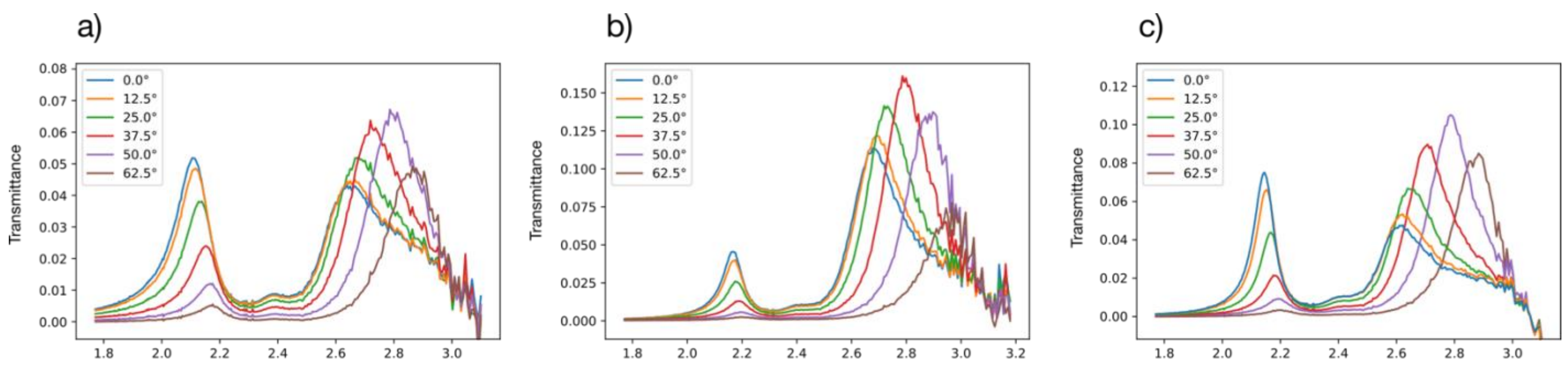

**Figure S4:** *Angle-resolved transmission spectra of three R6G:PVA microcavity samples with nominal dye loadings of 20%, 25%, and 33%, respectively.*

## Optical response of R6G in different chemical environments

The absorption spectrum of R6G strongly depends on its local chemical environment and aggregation state. Figure S5 compares the optical absorbance of R6G dissolved in ethanol, dispersed in a PVA:$H_2O$ solution, and embedded in a dried PVA thin film. In ethanol, R6G mainly exhibits the spectral features associated with a molecularly dissolved dye. In the aqueous PVA environment and, more prominently, in the dried polymer film, the absorption profile changes significantly. These spectral modifications are attributed to the formation of molecular aggregates. In particular, the emergence and enhancement of a blue-shifted shoulder are consistent with the presence of disordered H-type aggregates, while lower-energy contributions may be associated with J-like aggregate species.

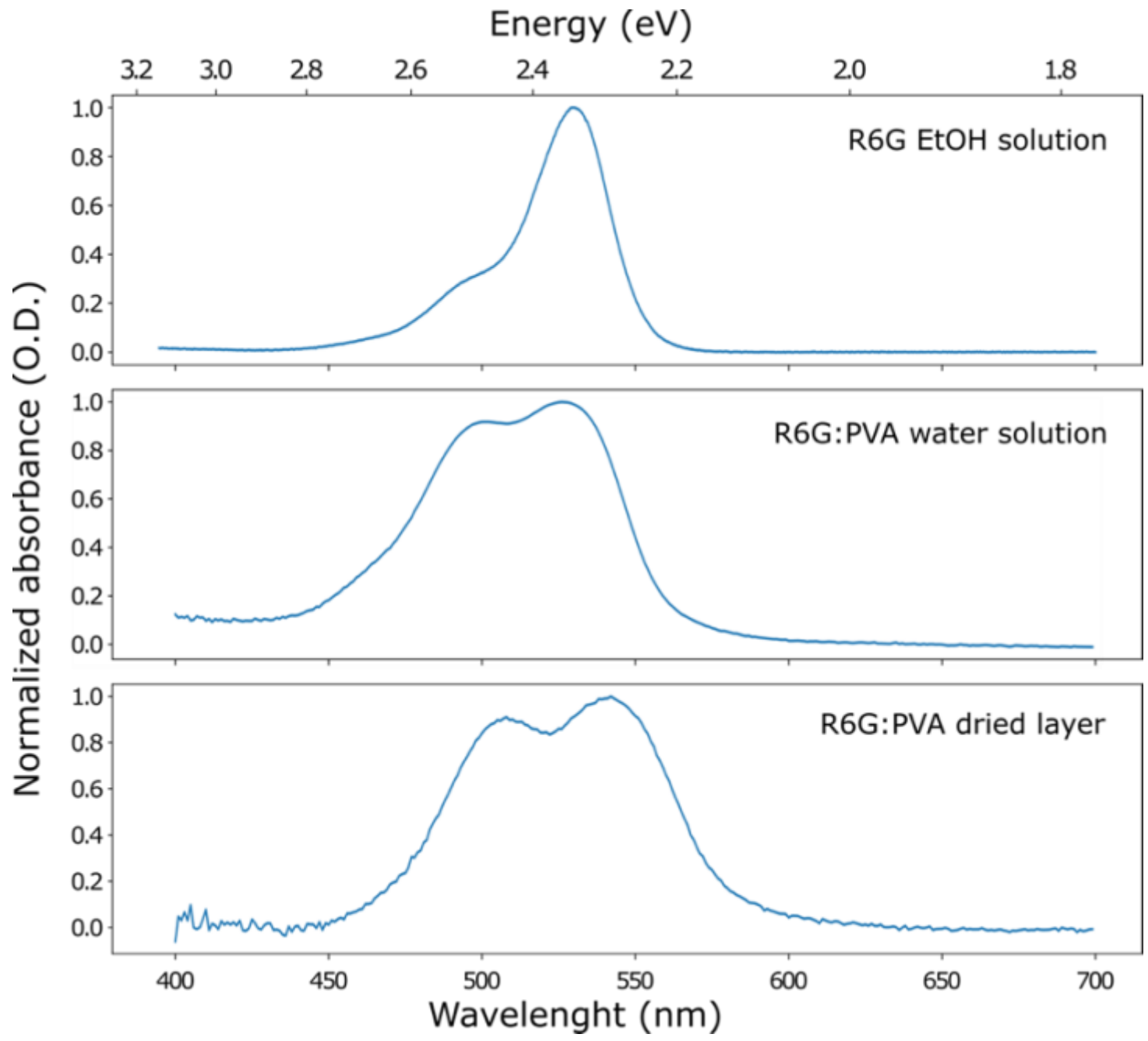

**Figure S5:** *Measured absorbance of R6G dissolved in ethanol (top), dispersed in a PVA:$H_2O$ solution (middle), and in a thin PVA-doped layer (bottom). The change of the profile due to aggregation is evident by passing from ethanol-based to water-based environments and in the dried film.*

**Extinction coefficient trends for varying cavity fabrication parameters**

The extinction coefficients, extracted by fitting the ellipsometric measurements (see Methods), were evaluated for R6G-doped PVA films prepared under different fabrication conditions. Figure S6a shows the extinction coefficient for samples with increasing R6G doping content in the PVA:$H_2O$ solutions, while keeping the spin-coating speed fixed at 65 rps. As the R6G concentration increases, the blue-shifted shoulder in the extinction spectrum becomes more pronounced. This trend indicates that higher dye loadings promote the formation of H-type aggregates. Therefore, the relative contribution of different aggregate species can be partially controlled through the dye concentration in the polymer matrix. The effect of spin-coating speed was also investigated at a fixed R6G concentration, as shown in Figure S6b. In this case, the extinction coefficient varies with the processing conditions, indicating that the final aggregation balance is affected not only by the chemical composition but also by the film formation dynamics. Changes in solvent evaporation rate, film thickness, and molecular packing during spin coating can modify the relative proportion of H- and J-type aggregates.

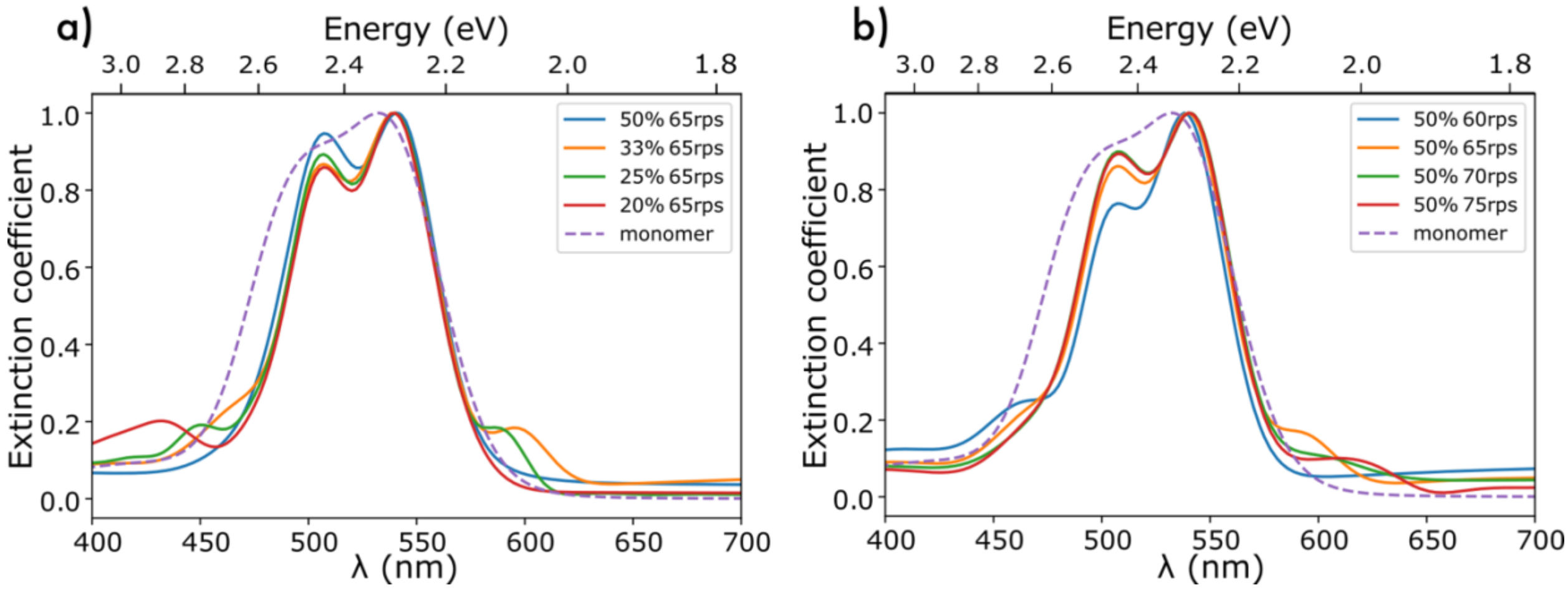

**Figure S6***: (a) Extinction coefficient of R6G-doped PVA films prepared with different R6G concentration in the PVA:$H_2O$ solution at a fixed spin-coating speed of 65 rps. The increasing intensity of the blue-shifted shoulder suggests enhanced H-aggregate formation at higher dye concentrations. (b) Extinction coefficient of R6G-doped PVA films prepared at different spin-coating speeds, from 60 to 75 rps, at fixed dye concentration. The observed spectral variations indicate that the aggregate distribution is sensitive to the fabrication parameters.*

**Middle polariton branch and double-exciton polariton model**

The polariton dispersion was initially analyzed using a single-exciton coupled-oscillator model based on the Tavis–Cummings Hamiltonian. Although anticrossing is clearly visible in the experimental transmission maps, the single-exciton model does not fully reproduce all the observed spectral features. This suggests that the molecular optical response cannot be adequately described by a single excitonic resonance. In addition, the presence of a partially resolved feature around 2.4 eV (Figure S4), consistent with the possible emergence of a middle-polariton branch, together with the double-peaked absorption spectrum shown in Figure 3, motivates the use of a double-exciton coupled-oscillator model. In this description, two nondispersive excitonic resonances are coupled to the angle-dependent cavity photon mode. The fitted polariton energies shown in Figure S7 were obtained by numerically diagonalizing, at each incidence angle, the following 3×3 Hamiltonian:

$$H=\begin{bmatrix} E_p & g_1 & g_2 \\ g_1 & E_{x1} & 0 \\ g_2 & 0 & E_{x2} \end{bmatrix};$$

The exciton energies $E_{x1}$ and $E_{x2}$ were constrained by the two absorption features observed in the corresponding uncapped R6G:PVA films. The effective refractive index $n_{eff}$ was treated as a free parameter within a physically reasonable range around the value extracted from the bare-cavity dispersion. The bare cavity energy at normal incidence, $E_0$, and the coupling strengths $g_1$ and $g_2$ were treated as free parameters.

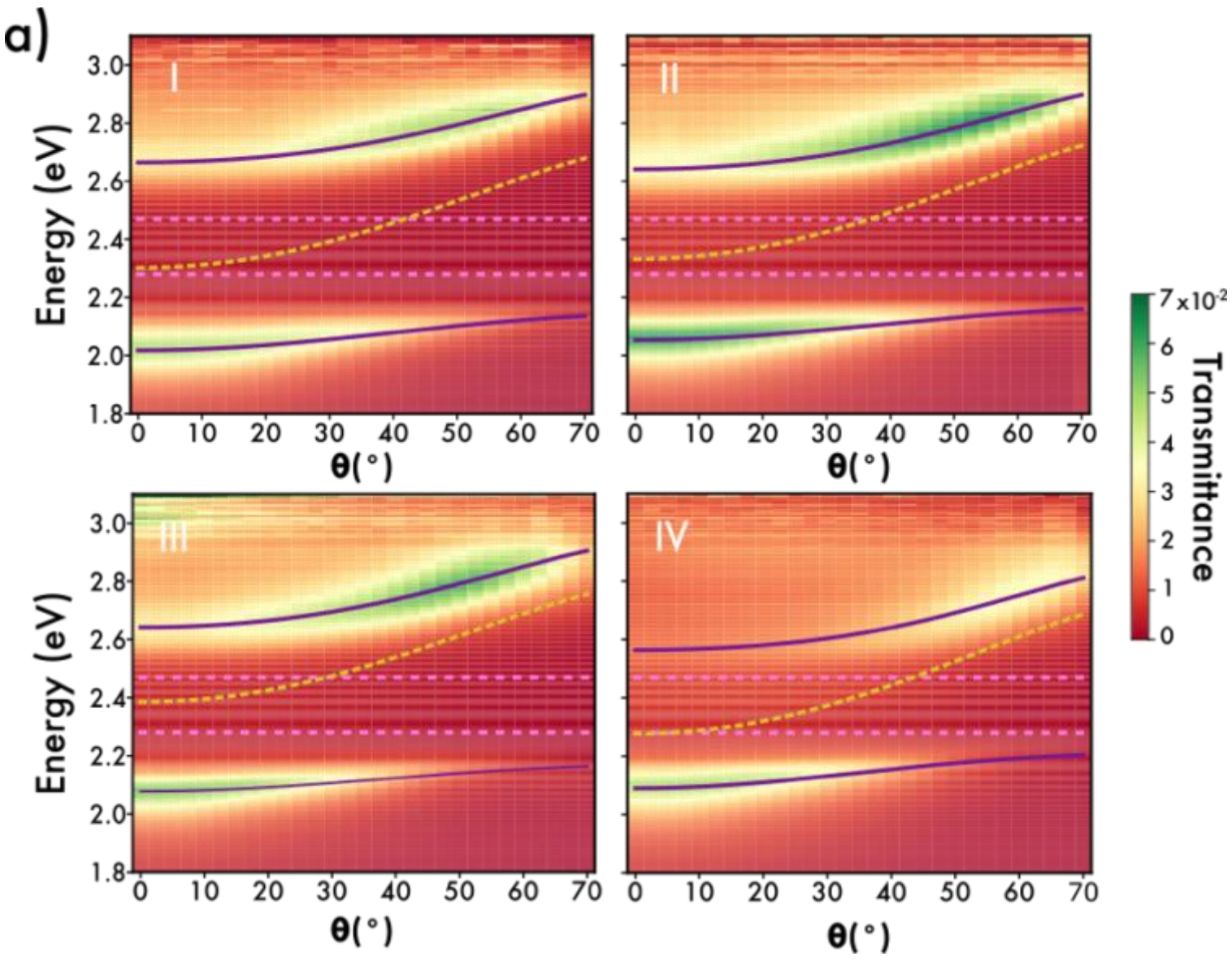

**Figure S7:** *Fit of the polariton dispersion using a double-exciton coupled-oscillator model for the same experimental data shown in Figure 2 of the main text. Two nondispersive excitonic transitions are coupled to the angle-dependent cavity photon mode. The model accounts for the simultaneous contribution of different R6G aggregate species to the observed polariton dispersion.*

### Aggregation as a function of the spin coating process

The dependence of the fitted cavity mode energy on the fabrication parameters was further analyzed for both the single- and double-exciton models. Figure S8a reports the cavity mode energy at normal incidence as a function of R6G concentration, while Figure S8b shows the corresponding trend as a function of spin-coating speed. The fitted cavity mode energies obtained from the single- and double-exciton models are in good agreement, indicating that the extracted bare cavity parameters are robust with respect to the choice of model, confirming that the main effect of the double-exciton model is to provide a more complete description of the molecular contribution.

**Figure S8:** *Fitted values of the cavity mode energy at normal incidence as a function of: (a) R6G concentration; (b) spin-coating speed. Black symbols correspond to the single-exciton model, while green symbols correspond to the double-exciton model. The agreement between*

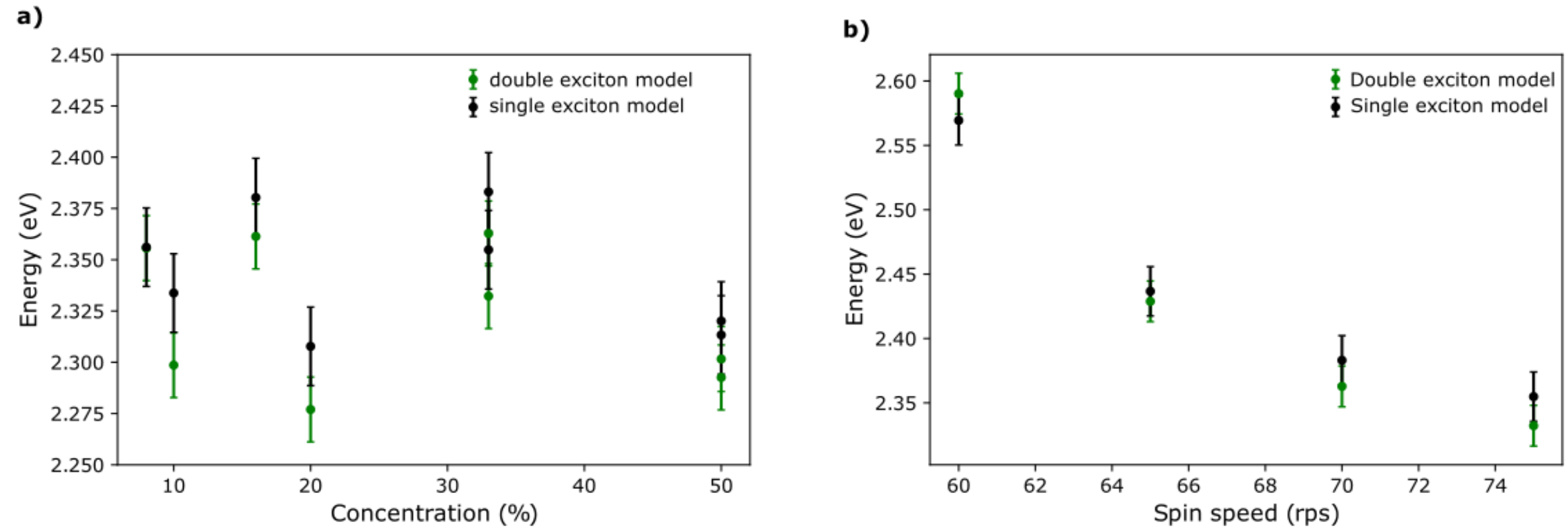

*the two models indicates that the extracted cavity energies are consistent across the fitting approaches.*

## Emission and excitation spectra

Emission and excitation measurements using a FLS1000 (Edinburgh Instruments) were performed to further characterize the optical response of the R6G-doped PVA films and to identify the molecular species contributing to polaritonic radiative recombination.

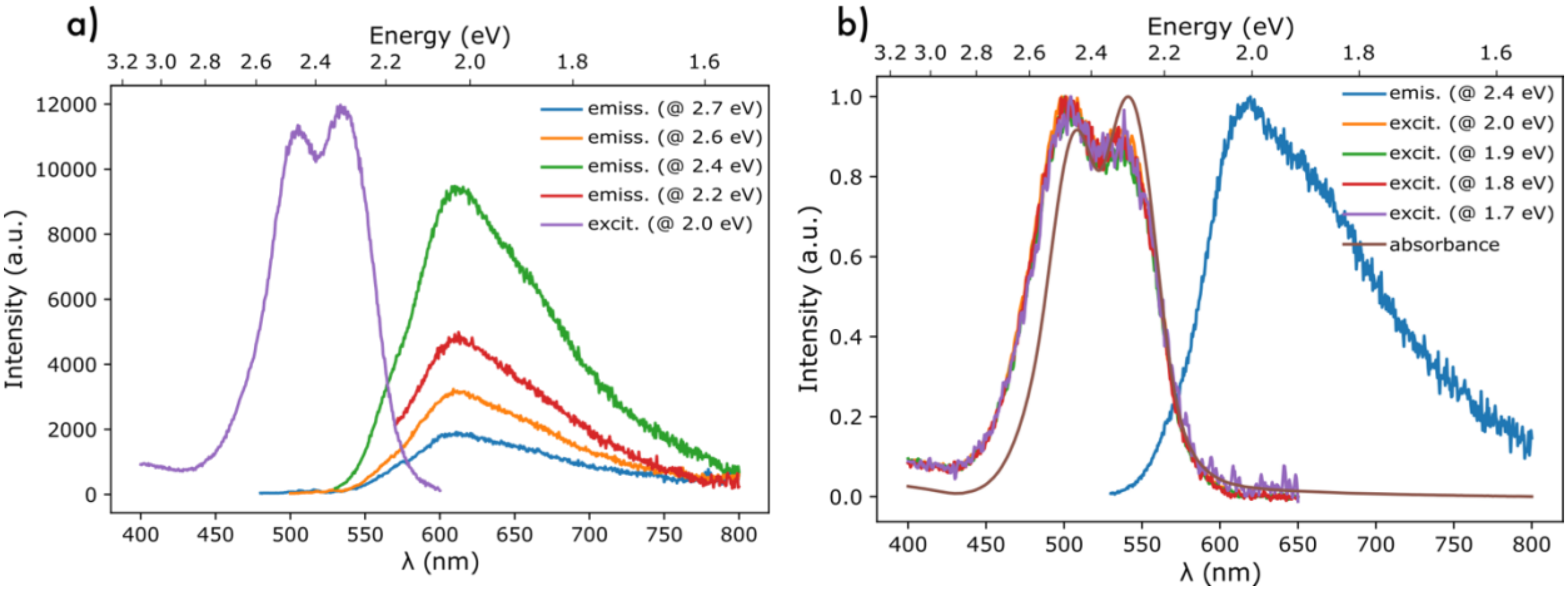

**Figure S9:** *a) Emission spectra of a 10% R6G:PVA thin film recorded at different excitation energies, together with the excitation spectrum monitored at the maximum of the emission band (2.0 eV, violet line). (b) Normalized emission spectrum of a 50% R6G-doped thin film (blue line), normalized excitation spectra monitored at the emission maximum (2.0 eV), at the shoulder (1.8 eV), and at nearby emission energies (1.7 and 1.9 eV), together with the normalized absorbance spectrum of the same film (brown line).*